\documentclass[letterpaper, 10 pt, conference]{ieeeconf}  

\IEEEoverridecommandlockouts                              
\usepackage{graphics} 
\usepackage{xcolor}
\usepackage{amssymb,latexsym,amsfonts,amsmath}
\usepackage{mathtools}
\usepackage{thmtools}

\usepackage{hyperref}
\hypersetup{hidelinks,linkcolor=black}

\usepackage{soul}
\setulcolor{red}

\declaretheorem{example}
\renewcommand\thmcontinues[1]{Continued}

\newcommand{\qed}{\hfill $\square$}

\newcommand{\A}{\mathcal{A}}
\newcommand{\B}{\mathcal{B}}
\newcommand{\W}{\mathcal{W}}
\newcommand{\CC}{\mathcal{C}}
\renewcommand{\S}{\mathcal{S}}
\newcommand{\U}{\mathcal{U}}
\newcommand{\G}{\mathcal{G}}
\newcommand{\scc}{\text{SCC}}
\newcommand{\R}{\mathbb{R}}
\newcommand{\N}{\mathbb{N}}
\newcommand{\Z}{\mathbb{Z}}
\newcommand{\inv}{\mathrm{inv}}
\newcommand{\intcc}[1]{\ensuremath{{\left[#1\right]}}}

\title{\LARGE \bf Numerical Estimation of Invariance Entropy for Nonlinear Control Systems}
\author{Mahendra Singh Tomar, Christoph Kawan, Pushpak Jagtap, and Majid Zamani
\thanks{This work is supported in part by the German Research Foundation (DFG) through the grants RU 2229/1-1, ZA 873/1-1, ZA 873/4-1, the H2020 ERC Starting Grant AutoCPS (grant agreement No.~804639), and the TUM International Graduate School of Science and Engineering (IGSSE).}
\thanks{M.~S.~Tomar and C.~Kawan are with the Computer Science Department, Ludwig Maximilian University of Munich, Germany. P.~Jagtap is with the Department of Electrical and Computer Engineering, Technical University of Munich, Germany. M.~Zamani is with the Computer Science Department, University of Colorado Boulder, USA, and with the Computer Science Department, Ludwig Maximilian University of Munich, Germany. Emails: {\tt\small mahendra.tomar@lmu.de}, {\tt\small christoph.kawan@lmu.de}, {\tt\small pushpak.jagtap@tum.de}, {\tt\small majid.zamani@colorado.edu}.}
}

\begin{document}

\maketitle
\thispagestyle{empty}
\pagestyle{empty}

\begin{abstract}
For a closed-loop control system with a digital channel between the sensor and the controller, the notion of \emph{invariance entropy} quantifies the smallest average rate of information transmission above which a given compact subset of the state space can be made invariant. In this work, we present for the first time an algorithm to numerically compute upper bounds of invariance entropy. With three examples, for which the exact value of the invariance entropy is known to us or can be estimated by other means, we demonstrate that the upper bound obtained by our algorithm is of the same order of magnitude as the actual value. Additionally, our algorithm provides a static coder-controller scheme corresponding to the obtained data-rate bound.
\end{abstract}

\section{Introduction}

In classical control theory, the sensors and controllers are usually connected through point-to-point wiring. In networked control systems (NCS), sensors and controllers are often spatially distributed and involve digital communication networks for data transfer.	Compared to classical control systems, NCS provide many advantages such as reduced wiring, low installation and maintenance costs, greater system flexibility and ease of modification. NCS find applications in many areas such as car automation, intelligent buildings, and transportation networks.	However, the use of communication networks in feedback control loops makes the analysis and design of NCS much more complex. In NCS, the use of digital channels for data transfer from sensors to controllers limits the amount of data that can be transferred per unit of time, due to the finite bandwidth of the channel. This introduces quantization errors that can adversely affect the control performance.%

The problem of control and state estimation over a digital communication channel with a limited bit rate has attracted a lot of interest in the past decade. In this context, a classical result, often called the \emph{data-rate theorem}, states that the minimal data rate or channel capacity above which a linear system can be stabilized or observed is given by the logarithm of the open-loop unstable determinant. This result has been proved under various assumptions on the system model, channel model, communication protocol, and stabilization/estimation objectives. Comprehensive reviews of results on data-rate-limited control can be found, e.g., in articles \cite{NairFagniniZampieriEvans07,andrievsky2010control,franceschetti2014elements} and books \cite{yuksel2013stochastic,matveev2009estimation,fang2017towards,kawan2013book}.%

For nonlinear systems, the smallest bit rate of a digital channel between the coder and the controller, to achieve some control task such as stabilization or invariance, can be characterized in terms of certain notions of \emph{entropy} which are described as intrinsic quantities of the open-loop system and are independent of the choice of the coder-controller. In spirit, they are similar to classical entropy notions used in the theory of dynamical systems to quantify the rate at which a system generates information about the initial state, see e.g.~\cite{katok1995introduction}.%

In this paper, we focus on a notion of \emph{invariance entropy} which was introduced in \cite{ColoniusKawan09} as a measure for the smallest average data rate above which a given compact and controlled invariant subset $Q$ of the state space can be made invariant. We present the first attempt to numerically compute upper bounds on the invariance entropy. Our approach combines different algorithms. First, we compute a symbolic abstraction of the given control system over the set $Q$ and the corresponding invariant controller. Particularly, we subdivide $Q$ into small boxes and assign control inputs (from a grid on the input set) to those boxes that guarantee invariance in one time step. This results in a typically huge look-up table whose entries are the pairs $(x,u)$ of states and control inputs which are admissible for maintaining invariance of $Q$. In the second step, the look-up table is significantly reduced by building a binary decision tree via a decision tree learning algorithm. This tree, in turn, leads to a typically smaller partition of $Q$ with one control input assigned to each partition element that will guarantee invariance of $Q$ in one time step. This data defines a map $T:Q \rightarrow Q$ to which, in the third step, we apply an algorithm that approximates the exponential growth rate of the total number of length-$n$ $T$-orbits which are distinguishable via the given partition. The output of this algorithm then serves as an upper bound for the invariance entropy.%

For the implementation of the first step --the construction of the invariant controller-- we use \texttt{SCOTS}, a software tool written in C++ designed for exactly this purpose \cite{rungger2016scots}. \texttt{SCOTS} relies on a rectangular grid, and assigns to each box in $Q$ a set of permissible control inputs. For the second step, we use the software tool \texttt{dtControl} \cite{ashok2020dtcontrol}, which builds the decision tree and determinizes the invariant controller by choosing from the set of permissible control inputs exactly one for each box. \texttt{dtControl} also groups together all the boxes which are assigned the same control input. For such a grouping, classification techniques such as logistic regression and linear support vector machines are employed. Finally, the third step is accomplished via an algorithm proposed in \cite{froyland2001rigorous}, originally designed for the estimation of topological entropy. This algorithm is based on the theory of symbolic dynamical systems and breaks up into standard graph-theoretic constructions.%
 
{\bf Brief literature review.} The notion of invariance entropy is equivalent to \emph{topological feedback entropy} that has been introduced earlier in \cite{NairEvansMarrelsMoran04}; see \cite{ColoniusKawanNair13} for a proof. Various offshoots of invariance entropy have been proposed to tackle different control problems or other classes of systems, see for instance \cite{Colonius12} (exponential stabilization), \cite{kawan2015network} (invariance in networks of systems), \cite{rungger2017invariance} (invariance for uncertain systems), \cite{colonius2018metric} (a measure-theoretic version of invariance entropy) and \cite{kawan2019invariance} (stochastic stabilization). Also the problem of state estimation over digital channels has been studied extensively by several groups of researchers. As it turns out, the classical notions of entropy used in dynamical systems, namely measure-theoretic and topological entropy (or small variations of them), can be used to describe the smallest data rate or channel capacity above which the state of an autonomous dynamical system can be estimated with an arbitrarily small error, see \cite{savkin2006analysis,liberzon2017entropy,sibai2017optimal,yang2018topological,kawan2018optimal}. Motivated by the observation that estimation schemes based on topological entropy suffer from a lack of robustness and are hard to implement, the authors of \cite{MaPo_automatica,PartII} introduce the much better behaved notion of \emph{restoration entropy} which characterizes the minimal data rate for so-called \emph{regular} and \emph{fine observability}. Finally, algorithms for state estimation over digital channels have been proposed in several works, in particular \cite{liberzon2017entropy,MaPo_automatica,hafstein2019numerical}.%

The paper is organized as follows. In Sect.~\ref{sec_prelim}, we introduce notation and the fundamental definitions. Section \ref{sec_implem} describes in details the implementation steps of our algorithm and illustrates them by a two-dimensional linear example. The results of our algorithm applied to one linear and two nonlinear examples are presented in Sect.~\ref{sec_ex}. Finally, Sect.~\ref{sec_conclusions} contains some comments on the performance of our algorithm and future works.%

\section{Notation and Preliminaries}\label{sec_prelim}

\subsection{Notation}

We denote by $\N, \Z, \Z_+$, and $\R$ the set of natural, integer, non-negative integer, and real numbers, respectively. For $a,b \in \Z$ with $a<b$, by $\intcc{a;b}$ we denote the set $\{i\in \Z\mid a\leq i\leq b\}$. By $[a_0a_1\ldots a_{N-1}]$, $a_i\in \N$ we denote a finite sequence of integers of length $N$, also called a \emph{word}. We use the notation $|\cdot|$ to denote the number of elements of a set, and also to denote the absolute value of a complex number. For an $n\times n$ matrix $B$, by $\lambda(B)$, $\rho(B)$ and $B_{i,j}$ we denote eigenvalues of $B$, the spectral radius and the entry in the $j$-th column of the $i$-th row, respectively.%

\subsection{Preliminaries}

Consider a discrete-time control system
\begin{equation}\label{eq:system}
  \Sigma:\quad x_{k+1} = f(x_k,u_k),%
\end{equation}
where $f:X\times U\to X$, $X\subseteq \R^n$, $U\subseteq \R^m$, is continuous. With $\U:=U^{\Z_+}$, let us define the transition map $\varphi:\Z_+\times X\times \U\to X$ of $\Sigma$ by%
\begin{equation*}
    \varphi(t,x,\omega) := \left\{ \begin{matrix}
                    x & \text{ if } t = 0, \\
                    f\big(\varphi(t-1,x,\omega),\omega(t-1)\big) & \text{ if } t\geq 1.
                \end{matrix} \right.%
\end{equation*}
We call a triple $(\A,\tau,G)$ an \emph{invariant partition} of $Q$, where $Q\subseteq X$, if $\A$ is a partition of $Q$, $\tau \in \N$, and $G:\A \to U^{\tau}$ is a map such that $\varphi(t,A,G(A))\subseteq Q$ for every $A\in \A$ and $t = 0,1,\ldots,\tau-1$.%

For a given $\CC = (\A,\tau,G)$, we define $T_{\CC}:Q\to Q$ as%
\begin{equation*}
  T_{\CC}(x) := \varphi(\tau-1,x,G(A_x)),%
\end{equation*} 
where $A_x\in \A$ is such that $x \in A_x$.%

{\bf{Invariance entropy:}} We call a set $Q$ \emph{controlled invariant} if for every $x\in Q$ there is $u\in U$ with $f(x,u) \in Q$. Let $Q$ be compact and controlled invariant. For $\tau\in\N$, a set $\S\subseteq \U$ is called $(\tau,Q)$-spanning if for every $x\in Q$ there exists $\omega\in\S$ such that $\varphi(t,x,\omega) \in Q$ for all $t\in \intcc{0;\tau-1}$. Let $r_{\inv}(\tau,Q)$ denote the number of elements in a minimal $(\tau,Q)$-spanning set. If there exists no finite $(\tau,Q)$-spanning set, we set $r_{\inv}(\tau,Q):=\infty$. Then the invariance entropy of $Q$ is defined as%
\begin{equation*}
  h_{\inv}(Q) := \lim_{\tau\to\infty}\frac{1}{\tau}\log_2 r_{\inv}(\tau,Q)%
\end{equation*}
if $r_{\inv}(\tau,Q)$ is finite for all $\tau>0$. Otherwise $h_{\inv}(Q):=\infty$. The existence of the limit follows from the subadditivity of the sequence $(\log_2 r_{\inv}(\tau,Q))_{\tau\in\N}$.%

{\bf{Counting entropy:}} Consider a set $Q$, a map $T:Q\to Q$ and a finite partition $\A = \{A_1,\ldots,A_q\}$ of $Q$. For $N \in \N$, consider the set $\W_N(T,\A) := \{[a_0a_1\ldots a_{N-1}]: \exists x\in Q \text{ with } T^i(x) \in A_{a_i}, 0\leq i < N\}$.%

The counting entropy of $T$ with respect to the partition $\A$ is defined as%
\begin{equation*}
  h^*(T,\A) := \lim_{N\to \infty}\frac{1}{N}\log_2|\W_N(T,\A)|,%
\end{equation*}
where the existence of the limit follows from the subadditivity of the sequence $(\log_2|\W_N(T,\A)|)_{N\in\N}$. Then the invariance entropy of $Q$ satisfies \cite[Thm.~2.3]{kawan2013book}%
\begin{equation*}
  h_{\inv}(Q) = \inf_{\CC = (\A,\tau,G)}\frac{1}{\tau}h^*(T_{\CC},\A),%
\end{equation*}
where the infimum is taken over all invariant partitions $\CC = (\A,\tau,G)$ of $Q$.%

To find an upper bound of $h^*(T_{\CC},\A)$, we select a refinement $\B = \{B_1,\ldots,B_{\bar{n}}\}$ of $\A$, i.e., $\B$ is a partition of $Q$ such that each element of $\A$ is the union of some elements of $\B$. Let us define an $\bar{n}\times \bar{n}$ \emph{transition matrix} $B$ by%
\begin{equation}\label{eq:transitionMatrix}
   B_{i,j} := \left\{\begin{matrix}
                      1& \text{ if } T_{\CC}(B_i)\cap B_j \neq \emptyset, \\ 
                      0& \text{ otherwise.}
                    \end{matrix}\right.%
\end{equation}
A sequence $[b_0b_1\ldots b_{N-1}]$ is called a $\B$-word if $B_{b_i,b_{i+1}} = 1$ for every $i\in\intcc{0;N-2}$. Next, we define the set%
\begin{align*}
  \W_N(\B,\A) &:= \big\{[a_0\ldots a_{N-1}]\mid \exists \text{ a } \B \text{-word } [b_0\ldots b_{N-1}] \\
	&\qquad\qquad\qquad\quad \text{s.t. } B_{b_i}\subseteq A_{a_i}, i \in \intcc{0;N-1}\big\}.%
\end{align*}
From \cite[Sec.~2.2]{froyland2001rigorous}, we have%
\begin{equation*}
  h(\B,\A) := \lim_{N\to\infty}\frac{\log_2|\W_N(\B,\A)|}{N} \geq h^*(T_{\CC},\A).%
\end{equation*}
Moreover, under certain assumptions it can be shown that $h(\B,\A)$ converges to $h^*(T_{\CC},\A)$ as the maximal diameter of the elements of $\B$ tends to zero, see \cite[Thm.~4]{froyland2001rigorous}.%

To compute $h(\B,\A)$, we first construct a directed graph $\G$ from the transition matrix $B$.%

We define a map $L:\B \to \{1,\ldots,|\A|\}$ by%
\begin{equation}\label{eq:labelMap}
  L(B_i) := j, \mbox{\quad where } j \mbox{ satisfies } B_i \subset A_j%
\end{equation} 
and call $L(B_i)$ the \emph{label} of $B_i$.%

The graph $\G$ has $\B$ as its set of nodes. If $B_{i,j}=1$, $i,j\in[1;\bar{n}]$ then there is a directed edge from the node $B_i$ to $B_j$ with the edge label $L(B_i)$. Elements of $\W_N(\B,\A)$ are generated by concatenating labels along walks of length $N$ on the graph $\G$.
Next we construct a second graph ${\G}_R$ which is deterministic (i.e., no two outgoing edges have the same label) and is such that the set of all bi-infinite words that are generated by walks on $\G_R$ is the same as the set generated by walks on $\G$. For details on the construction of $\G_R$ from $\G$, see \cite[Sec.~2.4]{froyland2001rigorous}. Each node in the deterministic graph $\G_R$ denotes a subset of $\B$ and has at most one outgoing edge for any given label. We use the graph $\G_R$ to define an \emph{adjacency matrix} $R$ by $R_{i,j} := l$, where $i,j\in[1;\tilde{n}]$ and $l$ is the number of edges from the node $i$ to the node $j$ of $\G_R$ and $\tilde{n}$ is the number of nodes in $\G_R$. If $\G$ is strongly connected, then from \cite[Prop.~7]{froyland2001rigorous}, we have $h(\B,\A) = \log_2 \rho(R)$. Thus, we have an upper bound for the invariance entropy of $Q$:%
\begin{equation*}
  h_{\inv}(Q) \leq \frac{1}{\tau}\log_2 \rho(R).%
\end{equation*}
If $\G$ is not strongly connected, we need to determine its strongly connected components and apply the algorithm separately to each component. Then the maximum of the specral radii of the obtained adjacency matrices will serve as an upper bound for $h_{\inv}(Q)$. Throughout this paper, we only use invariant partitions with $\tau = 1$, and write $(\A,G)$ instead of $(\A,1,G)$.%

\section{Implementation}\label{sec_implem}

In this section, we present the algorithm used in this work to numerically compute an upper bound of the invariance entropy.
We illustrate the steps involved in the algorithm with the help of the following example.%

\begin{example}\label{ex:examp}
Consider the linear control system%
\begin{equation*}
  x_{k+1} = Ax_k + \left[\begin{array}{c} 1 \\ 1 \end{array}\right]u_k, \quad
     A = \left[\begin{array}{cc} 2 & 0 \\ 0 & \frac{1}{2} \end{array}\right]%
\end{equation*}
with $x_k \in \R^2$ and $u_k \in U= [-1,1]$. For the compact controlled invariant set $\bar{Q} = [-1,1] \times [-2,2]$, see \cite[Ex.~21]{colonius2019controllability}, we intend to compute an upper bound of the invariance entropy. \qed
\end{example}

Given a discrete-time system $\Sigma$ as in \eqref{eq:system} and a set $\bar{Q}\subseteq X$, we proceed according to the following steps:%

\begin{enumerate}
\item\label{st:getcontroller} Compute a symbolic invariant controller for the set $\bar{Q}$. Consider the hyperrectangle $\bar{Q}_X$ of smallest volume that encloses $\bar{Q}$. We use \texttt{SCOTS} to compute an invariant controller for $\Sigma$ with $\bar{Q}_X$ as the state space and $\eta_s$ and $\eta_i$ as the grid parameters for the state space and the input space, respectively.  Use of small $\eta_s$ results in a finer grid on the state space, i.e., a grid with boxes of smaller volumes, which generally results in a better upper bound. We denote the set of boxes in the domain of the computed controller by $\B$. Let $\B=\{B_1,\ldots,B_{\bar{n}}\}$ and $Q :=\cup_{B_i\in\B}B_i\subseteq \bar{Q}$.%
        
\begin{example}[continues=ex:examp]
We used \texttt{SCOTS} with $\bar{Q}_X=\bar{Q}$ as the state space and the state space and input space grid parameters $\eta_s=0.57142$ and $\eta_i=0.005$, respectively. This results in a state space grid with $21$ boxes, $\B=\{B_1,\ldots,B_{21}\}$ and $Q=\bar{Q}$ (see Fig.~\ref{fg:linear2D:grid}). \qed
\end{example}
     
\item The controllers obtained in the previous step are in general non-deterministic, thus in this step, we determinize the obtained controller. We denote the closed-loop system ($\Sigma$ with the determinized controller $C$) by $\Sigma_C$. To determinize the controller efficiently, one can use the state-of-the-art toolbox \texttt{dtControl} \cite{ashok2020dtcontrol}, which utilizes the \emph{decision tree learning algorithm} to provide different determinized controllers with various choices of the input arguments `Classifier' and `Determinizer'. The tool not only determinizes the controller but also provides the required coarse partition $\mathcal{A}$ (of which $\B$ is a refinement). We refer the interested reader to \cite{ashok2020dtcontrol} for a detailed discussion about \texttt{dtControl}.%
    
\begin{example}[continues=ex:examp]
For this example, we used \texttt{dtControl} with parameters, Classifier = `cart' and Determinizer = `maxfreq'. This results in an invariant partition $(\A,G)$ for the set $Q:=\cup_{B\in\B}B$, where $\A$ is a partition of $Q$ such that every $A\in\A$ is a union of the members of some subset of $\B$ and $G(A)\in U$ is the control input assigned to the set $A$ given by \texttt{dtControl}. Figure~\ref{fg:linear2D:grid} shows the obtained partitions $\A$ and $\B$. \qed
\end{example}
		   
\begin{figure}[tp]
	\centering
	\includegraphics[scale=.6]{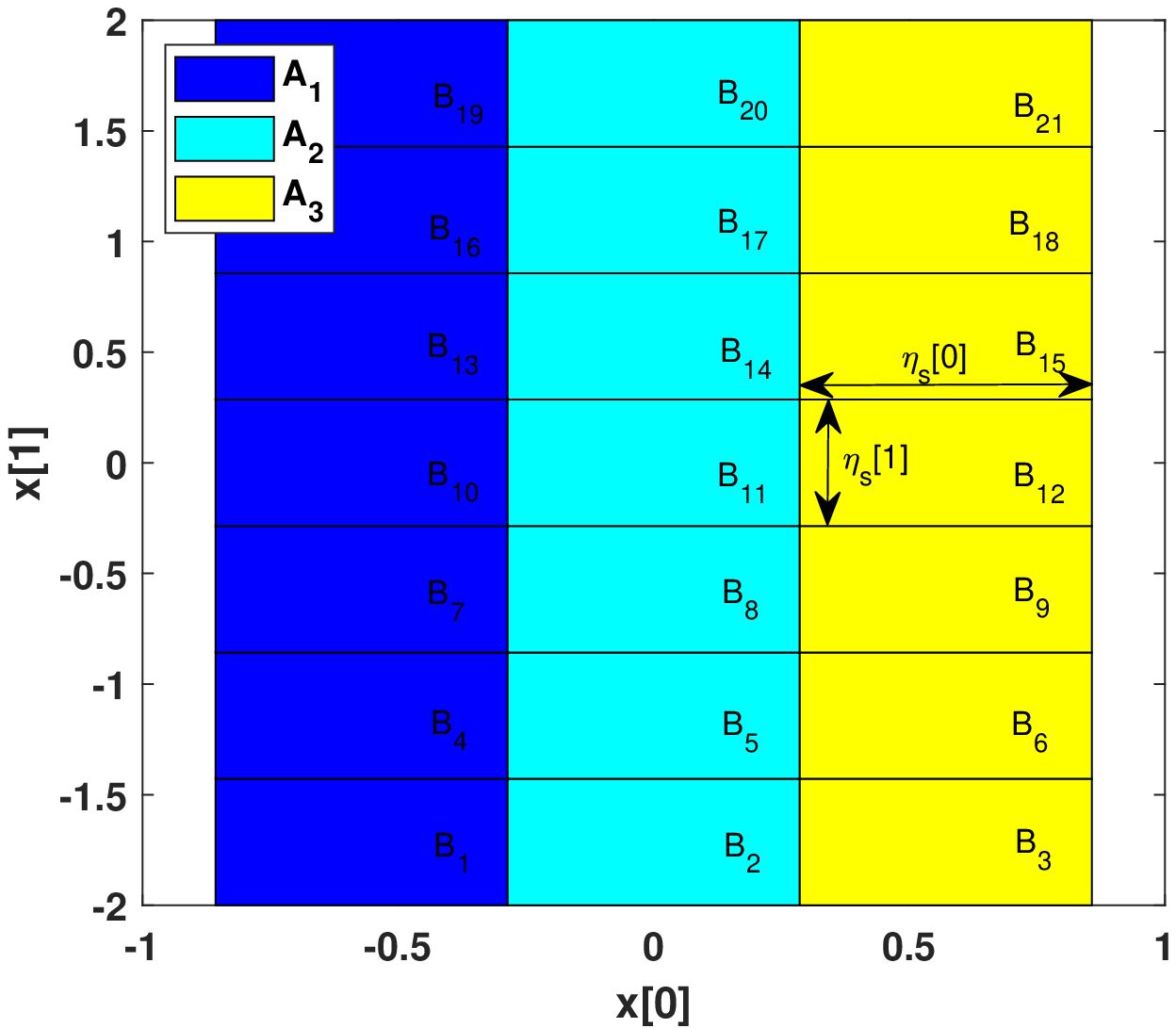}
	\caption{The partitions $\A=\{A_1,A_2,A_3\}$ and $\B=\{B_1,\ldots,B_{21}\}$ for Example \ref{ex:examp}.}
	\label{fg:linear2D:grid}
\end{figure}
    
\item For the dynamical system $\Sigma_C$, obtain the transition matrix $B$ \eqref{eq:transitionMatrix} for the boxes in $Q$.%

\begin{example}[continues=ex:examp]
$B$ comes out to be a $21\times 21$ matrix where each entry takes value $0$ or $1$ according to \eqref{eq:transitionMatrix}. \qed
\end{example}

\item Obtain the map $L$ as given in \eqref{eq:labelMap} that assigns a label to every member of the partition $\B$.
\begin{example}[continues=ex:examp]
        For $B_i\in \B$ 
        \begin{align*}
            L(B_i) = \left\{ \begin{matrix}
            1 & \text{ if } i=1+3t, 0\leq t\leq 6\\
            2 & \text{ if } i=2+3t, 0\leq t\leq 6\\
            3 & \text{ if } i=3+3t, 0\leq t\leq 6
            \end{matrix}
            \right. 
        \end{align*} 
        \qed
\end{example}
\item Construct a directed graph $\G$ with $\B$ as the set of nodes. If $B_{i,j} = 1$, then there is a directed edge from the node $B_i$ to $B_j$ with label $L(B_i)$.
\begin{example}[continues=ex:examp]
Figure~\ref{fg:linear2D:Gdigraph} shows the obtained directed graph $\G$. \qed
\end{example}
    
\begin{figure}[tp]
\centering
	\includegraphics[scale=.5]{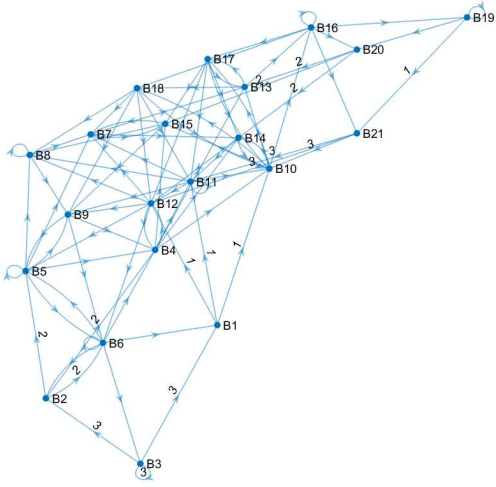}
	\caption{The directed graph $\G$ for Example \ref{ex:examp}. Some of the edge labels are omitted for clarity of the figure.}
\label{fg:linear2D:Gdigraph}
\end{figure}

\item Obtain the set $\G_\scc := \{\G_{\scc,1},\dots,\G_{\scc,p}\}$, where $\G_{\scc,i}$, $1\leq i\leq p$ is a strongly connected component of the graph $\G$. A directed graph is called strongly connected if for every pair of nodes $u$ and $v$ there exists a directed path from $u$ to $v$ and vice versa.

\begin{example}[continues=ex:examp]
$\G$ is strongly connected. Thus, $\G_{\scc} = \{\G\}$. \qed
\end{example}
    
\item For every $\bar{\G}\in \G_\scc$, find an associated deterministic graph $\bar{\G}_R$. The directed graph $\bar{\G}_R$ is deterministic in the sense that for every node no two outgoing edges have the same label.
   
\begin{example}[continues=ex:examp]
Figure~\ref{fg:linear2D:gccdigraph} shows the directed graph $\bar{\G}_R$.  Each node in Fig.~\ref{fg:linear2D:gccdigraph} refers to a subset of $\B$: $R_1=\{B_i\mid 13\leq i\leq 18\}$, $R_2=\{B_i\mid 7\leq i\leq 12\}$, $R_3=\{B_i\mid 4\leq i\leq 9\}$, $R_4=\{B_i\mid 16\leq i\leq 21\}$, $R_5=\{B_i\mid 10\leq i\leq 15\}$, $R_6=\{B_i\mid 1\leq i\leq 6\}$. \qed
\end{example}
    
\begin{figure}[tp]
	\centering
	\includegraphics[scale=.5]{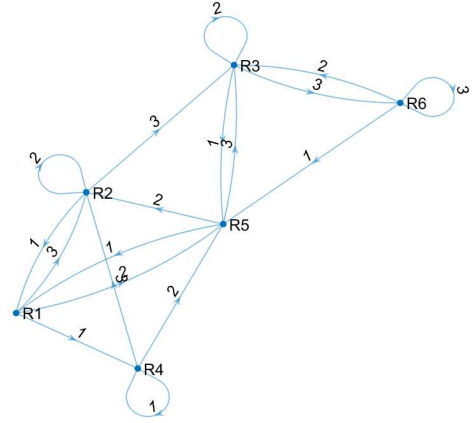}
\caption{The deterministic directed graph $\bar{\G}_R$ for Example~\ref{ex:examp}.}
	\label{fg:linear2D:gccdigraph}
\end{figure}
    
\item Using $\bar{\G}_R$, construct an adjacency matrix $R^{\bar{\mathcal{G}}}$ with%
\begin{equation*}
  R^{\bar{\G}}_{i,j} = l,%
\end{equation*}
where $l$ is the number of edges from node $i$ to node $j$ in $\bar{\G}_R$. Then we obtain%
\begin{equation*}
  h(\B,\A) = \max_{\bar{\G}\in \G_\scc} \log_2 \rho(R^{\bar{\G}}).%
\end{equation*}
        
\begin{example}[continues=ex:examp]
From $\bar{\G}_R$ we get%
\begin{equation*}
          R^{\bar{\G}} = \left[       \begin{matrix}
               0 & 1 & 0 & 1 & 1 & 0 \\
               1 & 1 & 1 & 0 & 0 & 0 \\
               0 & 0 & 1 & 0 & 1 & 1 \\
               0 & 1 & 0 & 1 & 1 & 0 \\
               1 & 1 & 1 & 0 & 0 & 0 \\
               0 & 0 & 1 & 0 & 1 & 1
            \end{matrix}\right],%
            \end{equation*}
             $\rho(R^{\bar{\G}}) = 3$ and $h(\B,\A) \approx 1.5850$. \qed
\end{example}
\end{enumerate}

\section{Examples}\label{sec_ex}

In the first two examples, we use known formulas for the invariance entropy, which have been proved for versions of invariance entropy that slightly differ from the one we introduced in Section \ref{sec_prelim}. However, from a numerical point of view, this should not make a considerable difference. In any case, the claimed values for $h_{\inv}(\bar{Q})$ in both cases are theoretical lower bounds, while our algorithm provides upper bounds.%

\subsection{A linear discrete-time system}

\begin{example}[continues=ex:examp]
Again consider the linear control system and the set $\bar{Q}$ as in the preceding section. The invariance entropy of $\bar{Q}$ is given by%
\begin{equation*}
  h_{\inv}(\bar{Q}) = \sum_{|\lambda(A)|\geq 1}\log_2|\lambda(A)| = 1.%
\end{equation*}
Table \ref{tb:linear2D:CandD} lists the obtained upper bounds $h(\B,\A)$ of $h_{\inv}(\bar{Q})$ with \texttt{SCOTS} parameters $\eta_s=0.01$ and $\eta_i=0.5$, for different choices of options in \texttt{dtControl}. \qed
\end{example}

\begin{table}
\centering
\caption{{Upper bound $h(\B,\A)$ for Example \ref{ex:examp} with different choices of the determinization options in \texttt{dtControl}. $h_{\inv}(\bar{Q}) = 1$}}
\begin{tabular}{|l|l|l|l|}
\hline
Classifier  & Determinizer & $|\A|$ & $h(\B,\A)$     \\ \hline\hline
cart              & maxfreq      & 4                                                                      & 1.0149 \\ \hline
logreg            & maxfreq      & 4                                                                      & 1.0149 \\ \hline
cart              & minnorm      & 5                                                                      & 1.0517 \\ \hline
logreg            & minnorm      & 5                                                                      & 1.0517 \\ \hline
\end{tabular}
\label{tb:linear2D:CandD}
\end{table}

\subsection{A scalar continuous-time nonlinear control system}

\begin{example}\label{ex:scalarNonlinear}
Consider the following scalar continuous-time control system discussed in \cite[Ex.~7.2]{kawan2013book}:%
\begin{equation*}
  \Sigma:\quad \dot{x} = (-2b\sin x \cos x - \sin^2 x + \cos^2 x) + u\cos^2 x,%
\end{equation*}
where $u \in [-\rho,\rho]$, $b>0$ and $0 < \rho < b^2 + 1$. The equation describes the projectivized linearization of a controlled damped mathematical pendulum at the unstable position, where the control acts as a reset force.%

The following set is controlled invariant:%
\begin{align*}
  & \bar{Q} = \\
	&\Bigl[\arctan (-b - \sqrt{b^2 + 1 + \rho} ),\arctan(-b - \sqrt{b^2 + 1 - \rho}) \Bigr].%
\end{align*}
In fact, $\bar{Q}$ is the closure of a control set, i.e., a maximal set of complete approximate controllability. With $\tau\in\R_{>0}$ as the sampling time, we first obtain a discrete-time system as in \eqref{eq:system}. Theory suggests that the following formula holds for the invariance entropy of $\Sigma$, see\footnote{The factor $\ln(2)$ appears due to the choice of the base-$2$ logarithm instead of the natural logarithm, which is typically used for continuous-time systems.}~\cite[Ex.~7.2]{kawan2013book}:%
\begin{equation*}
  h_{\inv}(\bar{Q}) = \frac{2}{\ln2}\sqrt{b^2 + 1 - \rho}.%
\end{equation*}
Discretizing the continuous-time system with sampling time $\tau$ results in a discrete-time system $\Sigma^{\tau}$ that satisfies%
\begin{equation*}
  h_{\inv}(\bar{Q};\Sigma^{\tau}) \geq \tau \cdot h_{\inv}(\bar{Q}) = \frac{2\tau}{\ln2}\sqrt{b^2 + 1 - \rho}.%
\end{equation*}
The inequality is due to the fact that continuous-time open-loop control functions are lost due to the sampling (since only the piecewise constant control functions, constant on each interval of the form $[k\tau,(k+1)\tau)$, $k \in \Z_+$, are preserved under sampling). Table \ref{tb:scalarNonlinear:tau:rho1b1} and \ref{tb:scalarNonlinear:tau:rho50b10} list the values of $h(\B,\A)$ for different choices of the sampling time with the parameters ($\rho=1$, $b=1$, $\eta_s=10^{-5}$, $\eta_i=0.2\rho$) and ($\rho=50$, $b=10$, $\eta_s=10^{-6}$, $\eta_i=0.2\rho$), respectively. For both of the tables, the \texttt{dtControl} parameters are Classifier = `cart' and Determinizer = `maxfreq'. Table \ref{tb:scalarNonlinear:CandD} shows the values of $h(\B,\A)$ for different selections of the coarse partition $\A$ with the parameters $\tau=0.01$, $\eta_s=10^{-6}$, $\eta_i=0.2\rho$, $\rho=1$, $b=1$. \qed
\end{example}

\begin{table}[]
\centering
\caption{Values of $h(\B,\A)$ and the invariance entropy $h_{\inv}(\bar{Q})$ for example~\ref{ex:scalarNonlinear} with $\rho=1$, $b=1$ and different choices of the sampling time $\tau$. $h_{\inv}(\bar{Q}) = 2.8854$}
\begin{tabular}{|l|l|l|}
    \hline
$\tau$ & $|\A|$ & $h(\B,\A)/\tau$  \\  \hline\hline
0.8    & 11            & 4.0207                       \\ \hline
0.5    & 6             & 4.0847                       \\ \hline
0.1    & 2             & 4.744                        \\ \hline
0.01   & 2             & 5.1994                       \\ \hline
0.001  & 2             & 24.7                         \\ \hline
\end{tabular}
\label{tb:scalarNonlinear:tau:rho1b1}
\end{table}

\begin{table}[]
\centering
\caption{Values of $h(\B,\A)$ and the invariance entropy $h_{\inv}(\bar{Q})$ for example~\ref{ex:scalarNonlinear} with $\rho=50$, $b=10$ and different choices of the sampling time $\tau$. $h_{\inv}(\bar{Q}) = 20.6058$}
\begin{tabular}{|l|l|l|}
        \hline
$\tau$ & $|\A|$ & $h(\B,\A)/\tau$ \\ \hline\hline
0.11   & 15            & 28.5012                     \\ \hline
0.1    & 11            & 29.1723                     \\ \hline
0.01   & 2             & 34.4707                   \\ \hline
0.001  & 2             & 55.5067                    \\ \hline
0.0001 & 2             & 1.5635e+03                 \\ \hline

\end{tabular}
\label{tb:scalarNonlinear:tau:rho50b10}
\end{table}

\begin{table}[]
\centering
\caption{Values of $h(\B,\A)$ for Example \ref{ex:scalarNonlinear} with
different choices of \texttt{dtControl} parameters. $h_{\inv}(\bar{Q})=2.8854$}
\begin{tabular}{|l|l|l|l|}
\hline
Classifier & Determinizer & $|\A|$ &   $h(\B,\A)/\tau$     \\ \hline\hline
cart       & maxfreq      & 2   & 5.1994     \\ \hline
logreg     & maxfreq      & 2   & 5.1994     \\ \hline
linsvm     & maxfreq      & 2   & 5.1994     \\ \hline
cart       & minnorm      & 11  & 6.4475     \\ \hline
logreg     & minnorm      & 11  & 6.4475     \\ \hline
\end{tabular}
\label{tb:scalarNonlinear:CandD}
\end{table}

\subsection{A 2d uniformly hyperbolic set}

\begin{figure}
	\centering
	\includegraphics[width=0.5\textwidth]{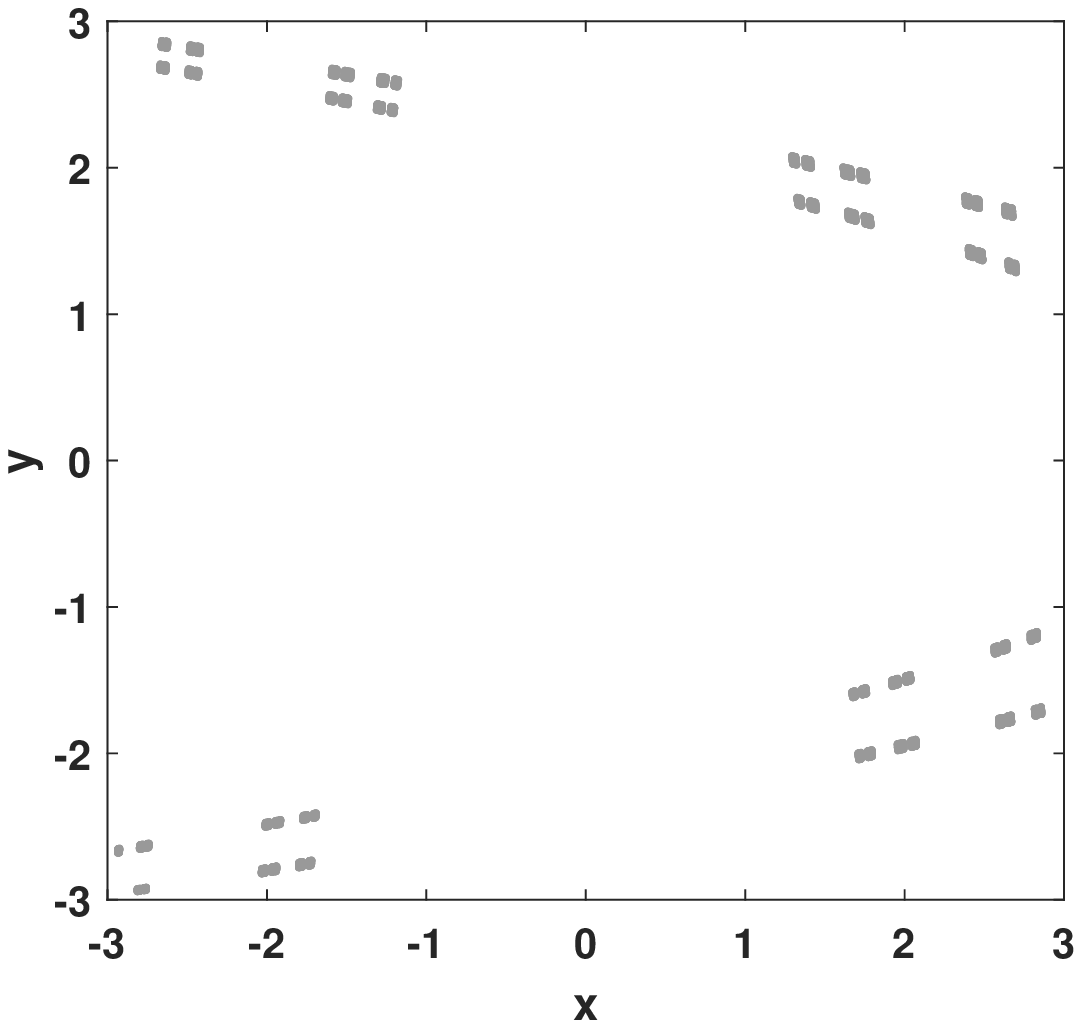}
	\caption{The domain of the controller in Example \ref{ex:hyperbolic} for invariance within the intersection of the original and time-reversed system}
	\label{fg:hyperbolic:domain}
\end{figure}

\begin{example}\label{ex:hyperbolic}
Consider the map%
\begin{equation*}
  f(x,y) := (5 - 0.3y - x^2,x),\quad f:\R^2 \rightarrow \R^2,%
\end{equation*}
which is a member of the famous and well-studied H\'enon family. We extend $f$ to a control system with additive control:%
\begin{equation*}
  \Sigma:\quad \left[\begin{array}{c} x_{k+1} \\ y_{k+1} \end{array}\right] = \left[\begin{array}{c} 5 - 0.3 y_k - x_k^2 + u_k \\ x_k + v_k \end{array}\right],%
\end{equation*}
where $\max\{|u_k|,|v_k|\} \leq \varepsilon$. It is known that $f$ has a non-attracting uniformly hyperbolic set $\Lambda$, which is a topological horseshoe (called the \emph{H\'enon horseshoe}). This set is contained in the square centered at the origin with side length \cite[Thm.~4.2]{robinson1998dynamical}%
\begin{equation*}
  r := 1.3 + \sqrt{(1.3)^2 + 20} \approx 5.9573.%
\end{equation*}
If the size $\varepsilon$ of the control range of $\Sigma$ is chosen small enough, the set $\Lambda$ is blown up to a compact controlled invariant set $Q^{\varepsilon}$ with nonempty interior which is not much larger than $\Lambda$; see \cite{kawan2020control}. Moreover, the theory suggests that as $\varepsilon\rightarrow0$, $h_{\inv}(Q^{\varepsilon})$ converges to the negative topological pressure of $f_{|\Lambda}$ with respect to the negative unstable log-determinant on $\Lambda$; see \cite{bowen2008equilibrium} for definitions. A numerical estimate for this quantity, obtained in \cite{froyland1999using} via Ulam's method, is $0.696$.%

We select $\bar{Q} = [-r/2,r/2]\times[-r/2,r/2]$.

For the parameter values $\varepsilon=0.009$, $\eta_s=0.0021$, $\eta_i=0.003$, the table \ref{tb:hyperbolic} lists the values of $h(\B,\A)$ for different selections of the coarse partition $\A$. To obtain the invariant controller whose domain is likely to approximate the (all-time controlled invariant) set $Q^{\varepsilon}$, we intersect the domain of the invariant controller in $\bar{Q}$ associated to the given system with that of its time-reversed system%
\begin{equation*}
  \Sigma^-:\quad \left[\begin{array}{c} x_{k+1} \\ y_{k+1} \end{array}\right] = \left[\begin{array}{c}  y_k - v_k \\ \frac{1}{0.3}(5 - (y_k - v_k)^2 + u_k - x_k) \end{array}\right].%
\end{equation*}
Figure \ref{fg:hyperbolic:domain} shows the intersection of the two domains. \qed
\end{example}

\begin{table}[]
\centering
\caption{Values of $h(\B,\A)$ for Example \ref{ex:hyperbolic} with different selections of \texttt{dtControl} options. $h_{\inv}(\bar{Q}) \approx 0.696$}
\begin{tabular}{|l|l|l|l|}
\hline
Classifier & Determinizer & $|\A|$   &   $h(\B,\A)$     \\ \hline\hline
cart       & maxfreq      & 1657  & 1.3178 \\ \hline
linsvm     & maxfreq      & 1656  & 1.3207 \\ \hline
cart       & minnorm      & 3608  & 1.7262 \\ \hline
logreg     & minnorm      & 2935  & 1.6927 \\ \hline
linsvm     & minnorm      & 3611  & 1.7255 \\ \hline
\end{tabular}
\label{tb:hyperbolic}
\end{table}

\addtolength{\textheight}{-3cm}   

\section{Conclusions}\label{sec_conclusions}

All the computations in this work were performed on an Intel Core i5-8250U processor with 8 GB RAM. The computation time and memory requirement of \texttt{SCOTS} increases with the reduction of the grid parameter values and the increase in the volume of the state and input set. For \texttt{dtControl}, the time increases with the size of the controller file obtained from \texttt{SCOTS}. The part of the implementation which computes the deterministic graph $\bar{\G}_R$ from the directed one $\bar{\G}$ is written as a MATLAB \texttt{mex} function. The computation time of the MATLAB code increases with the increase in the number of nodes and the number of edges in the graph $\bar{\G}$. For instance, Example \ref{ex:hyperbolic} needs 6.1264 GB of memory and a computation time of 11 min. A lower upper bound is expected with a reduction in the state space grid parameter $\eta_s$, but any value of $\eta_s$ less than 0.0021 in Example \ref{ex:hyperbolic} results in memory allocation issues on our machine.%

Future work will focus on the selection of better invariant partitions resulting in smaller upper bounds. For example, during determinization of the controller, for a given grid box $B$, those control input values are preferred which make a smaller set of grid boxes covering the image of $B$ under the system dynamics. Also, it is likely that by considering control sequences of length $\tau \geq 2$, rather than unit length, will result in better upper bounds.%




\bibliographystyle{IEEEtran}
\bibliography{numeric}

\end{document}